\documentclass[11pt, titlepage, reqno]{article}
 \usepackage[super]{natbib}
 \usepackage{amssymb, latexsym}
 \usepackage{amsmath} 
 \usepackage{amsthm}
 \usepackage{bm}
 \usepackage[mathscr]{eucal}
 \usepackage{enumerate}
 \usepackage{graphicx}
 \usepackage{float}

 \renewcommand{\section}[1]{\medskip \addtocounter{section}{1}\raggedright 
     \textbf{\Roman{section}. \ #1}\medskip \setcounter{subsection}{0}
    \setlength{\parindent}{5ex}
 }
 \renewcommand{\subsection}[1]{\medskip \addtocounter{subsection}{1}\raggedright
    \textbf{\Alph{subsection}. \ #1} \medskip \setcounter{subsubsection}{0}\setlength{\parindent}{5ex}
}

 \theoremstyle{plain}
\begin{document}

 \begin{titlepage}

 \begin{center}

 \huge{\textbf{Connecting the grain-shearing mechanism of wave propagation in marine sediments to fractional calculus}}\\
 
 

 \vspace{10ex}

\Large{Vikash Pandey}\footnote{e-mail: vikashp@ifi.uio.no\\ \textit {Preprint submitted to: The Journal of the Acoustical Society of America (JASA)  on December 11, 2015.}} and Sverre Holm

 Department of Informatics
 
 University of Oslo

 P.O. Box 1080, NO-0316 Oslo, Norway
 
 \end{center}

 \end{titlepage}

 \begin{abstract}

An analogy is drawn between the diffusion-wave equations derived from the fractional Kelvin-Voigt model and those obtained from Buckingham's grain-shearing (GS) model [J. Acoust. Soc. Am. \textbf{108}, 2796--2815 (2000)] of wave propagation in saturated, unconsolidated granular materials. The material impulse response function from the GS model is found to be similar to the power-law memory kernel which is inherent in the framework of fractional calculus. The compressional wave equation and shear wave equation derived from the GS model turn out to be the Kelvin-Voigt fractional-derivative wave equation and the fractional diffusion-wave equation respectively. Also, a physical interpretation of the characteristic fractional-order present in the Kelvin-Voigt fractional derivative wave equation and time-fractional diffusion-wave equation is inferred from the GS model. The shear wave equation from the GS model predicts both diffusion and wave propagation in the fractional framework. The overall goal is intended to show that fractional calculus is not just a mathematical framework which can be used to curve-fit the complex behavior of materials, but rather it can be justified from real physical process of grain-shearing as well.

 \end{abstract}

 \addtocounter{page}{2}

 \section{INTRODUCTION}

 \setlength{\parindent}{5ex}
 
It is well established that Biot's theory of wave propagation in fluid-saturated porous solids is inadequate for a wide class of earth materials, e.g. reservoir rock and marine sediments.\citep{Hovem1979}${}^-$\citep{Carcione2011} Biot's theory assumes a homogeneous rigid frame with fluid filled pores and predicts attenuation for compressional waves to vary quadratically in the low frequencies and to be constant in the high frequencies. When viscosity of the pore-fluid is taken into account, the theory predicts attenuation to rise with the square root of frequency for high frequencies, though the low frequency behavior remains unchanged. However, experimental observations from marine sediments demonstrate near-linear power-law dependency of attenuation on frequency above a characteristic frequency. Consequently, attempts have been made to improve Biot's theory and also new theories have been proposed.

\medskip

Probably, one of the first attempts for improvement came in the form of the Biot-Stoll theory which included two additional attenuation mechanisms into the original Biot's theory.\citep{Stoll1970} First, dissipation due to friction from the inter-granular sliding at the common contact surface, and second losses due to viscous dissipation caused by relative motion between the grains and the interstitial fluid. Although the model gives a good fit to the experimental measurements, causality is not ensured.\citep{Holland1988}${}^-$\citep{Chotiros1995}

\medskip

The next significant improvement is the addition of squirting mechanism into the classical Biot theory to form Biot/squirt (BISQ) theory.\citep{Dvorkin1994}${}^,$\citep{Dvorkin1995} The BISQ model assumes solid rock and introduces cracks and pores, such that squirting of the pore-fluid is in parallel as well as in the lateral direction when pores are compressed. The model bridges the attenuation from macroscopic quantities such as porosity, permeability, saturation, viscosity and compressibility, with the attenuation from the microscopic quantity of squirt-flow length. The BISQ theory resolved two issues which were not modeled in the classical Biot theory. The issues were; first, shift of relaxation towards the lower frequency for sediments saturated with viscous fluid, and second increase of compressional wave speed with pore-fluid viscosity. The BISQ theory is mathematically governed by a diffusion equation implying that the squirt-flow mechanism dominates the Biot's dissipative mechanism particularly in consolidated sediments with relatively low permeability.

\medskip

Now, we arrive at one of main themes of this article which is the role of grain-physics in fluid-saturated porous network.\citep{Diallo2000} Chotiros and Isakson\citep{Chotiros2004} developed the Biot--Stoll plus grain-contact squirt and shear flow (BICSQS) model which assumes Poiseuille flow to be valid for all frequencies and squirt flow to have any orientation. The response to the compressive and shear forces along the grain-contacts is modeled using viscoleastic models, namely the Zener model and the Kelvin-Voigt model respectively. The grain-contact squirt flow and grain-contact shear drag losses have less contribution in the low frequencies, but affect the result considerably in intermediate-to-high frequencies. Causality is ensured in this process and attenuation scales quadratically at low frequencies and near-linearly at high frequencies.

\medskip

The model investigated in this work is the alternative grain-shearing (GS) proposed by Buckingham for wave propagation in fluid-saturated, unconsolidated granular medium.\citep{Buckingham2000}${}^,$\citep{Buckingham2005} The development of the model can be traced back to the earlier works of Buckingham.\citep{Buckingham1997}${}^,$\citep{Buckingham1998} In comparison to the classical Biot theory, the GS model assumes the absence of an elastic frame and shows the generation of both compressional and shear waves from the strain-hardened intergranular sliding. In the last $15$ years, the GS model has undergone two improvements; first, in the form of the viscous grain shearing (VGS) model,\citep{Buckingham2007} and second, the recent VGS($\lambda$) model.\citep{Chotiros2010}${}^,$\citep{Buckingham2010} As the name implies, the VGS model takes into account the effect of pore-fluid viscosity on the GS process. Consequently, an additional relaxation time constant is included which helps in better fitting of experimental data at low frequencies. The GS and VGS models have different attenuation trends for low frequencies $\left(<10\mbox{ }kHz\right)$, but merge asymptotically at high frequencies. The assumption that the effect of pore-fluid viscosity is the same for both the compressional wave and the shear wave is modified in the VGS($\lambda$) model, which has two relaxation time constants, one each for the compressional and shear waves respectively. The parameter $\lambda$ denotes the wavelength dependence of fluid viscosity in damping of the waves.

\medskip

This paper which is an expanded version of\citep{Holm2016} builds on Buckingham's GS model and attempts to connect the physical mechanism of grain-shearing with the mathematical framework of fractional calculus. As we will show later in the paper, the wave equations obtained from the GS model can be mapped into the fractional framework to appear as Kelvin-Voigt fractional derivative diffusion-wave equations. The motivation behind this study is three-fold. First, the apparent good fit of the anomalous phase-velocity dispersion curve and power-law attenuation curve predicted from the GS model with the experimental data is also the main characteristic of materials modeled using fractional calculus. Although, the curve-fitting agreement is mostly in the range of $10$-$400$ kHz,\citep{Buckingham2007}${}^,$\citep{Buckingham2014} it provides enough impetus to study the GS model in the light of fractional calculus. In recent years, the methodology of fractional calculus has found extensive applications in the modeling of mechanical and wave dispersive properties of complex viscoelastic and poroelastic materials such as biological media\citep{Sebaa2006}${}^-$\citep{Zhang2016} and earth materials.\citep{Carcione2002}${}^-$\citep{Zhu2014} The similarity between the material impulse response function (MIRF) from the GS model and the power-law memory kernel of fractional calculus which we will later show in this paper suggests a strong connection between them. Second, the extensive mathematical utilities of fractional calculus provides a flexible but yet a robust framework for modeling of complex materials. Since, the framework is not constrained to integer-order derivatives, it predicts power-law attenuation of the wave $\alpha_{k}\propto \omega^{\gamma}$, where the exponent $\gamma$ can be any real positive number. This also facilitates the tracking of evolution of a given physical phenomenon into a different one; such as, from a damped diffusive process to a propagating wave.\citep{Agrawal2002}${}^,$\citep{Ray2007}

\medskip

Third, the parameter $\gamma$ which also corresponds to the order of the fractional wave equations is usually estimated by hit-and-trial choice of relaxation time ratios for curve-fitting of experimental data with the theoretically predicted curves.\citep{Horton1959}${}^,$\citep{Meidav1964} Although the fractional framework imparts greater flexibility to the fitting process, its application is restricted due to this uncertainty in order estimation. This limitation can be traced back to the way fractional-order wave equations are derived from adhoc phenomenological models comprising combinations of springs and dashpots.\citep{Mainardi2010} Consequently, a proper physical interpretation of the fractional-order $\gamma$ is still lacking. In this paper, we also aim to address this concern by relating the order with the physical parameters of the material. The correct knowledge of $\gamma$ would then reduce the ambiguities in the curve-fitting process. Moreover, without adopting a fractional calculus approach, such complex behavior could be difficult to model, both analytically as well as numerically.\citep{Holm2015} 

\medskip

The article is organized as follows. In Sec. II, we provide a short summary of Buckingham's GS model outlining its underlying physical mechanism. Then, in Sec. III, the framework of fractional calculus along with the Kelvin-Voigt fractional derivative wave equation and time-fractional diffusion-wave equation are introduced. The compressional and shear wave equations obtained from the GS model in Sec. II are then mapped into the domain of fractional calculus in Sec. IV. Also, the dispersive characteristics from the fractional wave equations as well as the emergence of squirt-flow mechanism from GS model are analyzed. Finally, in Sec. V, we discuss the implications of this work.

\medskip

\section{GRAIN-SHEARING MODEL}

Buckingham's\citep{Buckingham2000} GS model takes into account the non-Biot dissipative mechanisms of grain-shearing in saturated, unconsolidated marine sediments. Before proceeding further, it is worthwhile to mention that the GS mechanism has already been employed in the past to explain the anelastic behavior of metals and crystals. Zener\citep{Zener1941}${}^,$\citep{Zener1958} showed that slipping at grain boundaries reduces the effective elasticity of the bulk crystalline material. Also, in most cases, shearing at grain-boundaries is found more favourable than shearing of the individual grains themselves, which indeed corresponds to Buckingham's treatment.

\medskip

As illustrated in Fig. 1, the static overburden pressure develops micro-asperities at the contact surfaces of the sediment grains. The grains exhibit stick-slip motion triggered by the velocity gradient set up by the initial wave disturbance. The intergranular sliding is mediated through the saturating pore-fluid present between the grains. Buckingham attributes the strain-hardening mechanism to the viscous drag force created due to the sliding motion across the pore-fluid. The generated drag force would then increasingly oppose the motion with time. This time-dependency of the drag force is represented by a time-dependent viscous dashpot in the Maxwell element. As the grains slide along the radials of circle of contact, it gives rise to compressional and translational shearing which finally build up as compressional and shear waves respectively.

\begin{figure}[H]
\centering
\hspace*{0cm}
\includegraphics[scale=0.35]{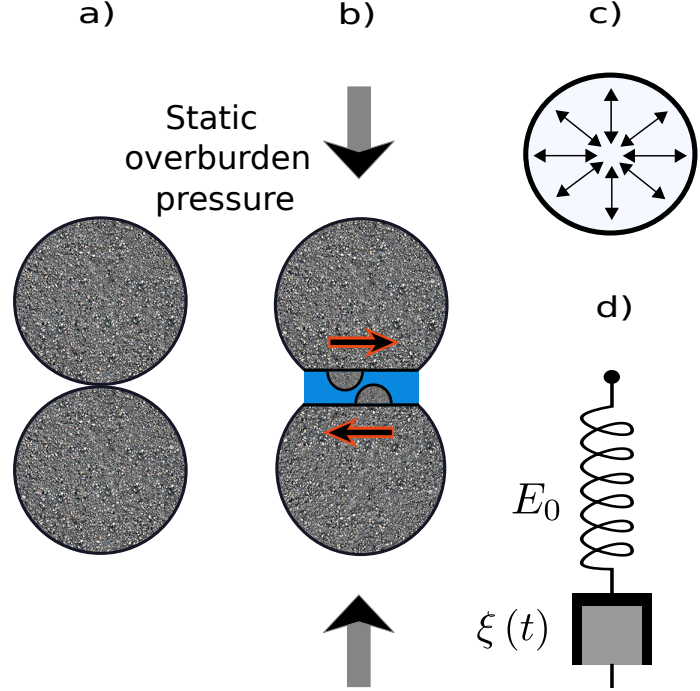}
\caption{(Color online) Grain-to-grain shearing mechanism: a) Two spherical, saturated mineral grains in light contact. b) Deformed grains due to the static overburden pressure, development of micro-asperities (small solid hemispheres) separated by a thin film of pore-fluid, and intergranular tangential and compressional shearing. c) Radials of the circle of contact of the grains. d) Equivalent modified Maxwell model consisting of a series combination of a Hookean spring $E_{0}$ and a time-dependent viscous dashpot $\xi\left(t\right)$. Figure adapted
from Buckingham.\citep{Buckingham2000}}
\end{figure}

\medskip

Buckingham writes the constitutive relation of the modified Maxwell model (see, Eqs. (13)--(16) in Buckingham\citep{Buckingham2000}) as
\begin{equation}\label{eq01}
\frac{1}{E_{0}}\frac{d\chi}{dt}+\frac{\chi}{\xi\left(t\right)}=\frac{d\varepsilon}{dt}
\end{equation}
where, $\chi$ is the stress, $\varepsilon$ is the strain, $E_{0}$ is the elastic modulus at zero frequency, t is the time and $\xi\left(t\right)$ is the time-dependent viscosity of the dashpot. Assuming linear wave propagation, the viscosity $\xi\left(t\right)$ is approximated as 
\begin{equation}\label{eq02}
\xi\left(t\right)\approx\xi_{0}+\theta t
\end{equation}
where the zero-order term $\xi_{0}$ represents the viscosity of the pore-fluid before the sliding is triggered and the first-order term $\theta=\frac{d\xi}{dt}\bigr|_{t=0}\geq0$ is the strain-hardening coefficient. The stress relaxation for the Maxwell element is derived as (see, Eqs. (18) and (19) in Buckingham\citep{Buckingham2000})
\begin{equation}\label{eq03}
\chi=\chi_{0}\left(1+\frac{\theta}{\xi_{0}}t\right)^{-\frac{E}{\theta}},\mbox{ where }\chi_{0}=\left|\varepsilon\right|E_{0}.
\end{equation}
The time-dependent term of Eq. (\ref{eq03}) along with the introduction of a leading term is identified as the pulse shape function $h\left(t\right)$ and written as:
\begin{equation}\label{eq04}
h\left(t\right)=\frac{\theta}{\xi_{0}}\left(1+\frac{\theta}{\xi_{0}}t\right)^{-\frac{E}{\theta}}.
\end{equation}
Since the stick-slip processes are randomly distributed in the medium, they are therefore ensemble averaged to obtain the material impulse response functions (MIRFs) $h_{p}\left(t\right)$ and $h_{s}\left(t\right)$ for compressional waves and shear waves respectively. The MIRFs are given as (see, Eqs. (25)--(28) in Buckingham\citep{Buckingham2000}):
\begin{equation}\label{eq05}
h_{p}\left(t\right)=t_{p}^{-1}\left(1+\frac{t}{t_{p}}\right)^{-\gamma_{p}}\text{, where \ensuremath{t_{p}=\frac{\xi_{0p}}{\theta_{p}}} and \ensuremath{\gamma_{p}=\frac{E_{0p}}{\theta_{p}}}}
\end{equation}
and
\begin{equation}\label{eq06}
h_{s}\left(t\right)=t_{s}^{-1}\left(1+\frac{t}{t_{s}}\right)^{-\gamma_{s}}\text{, where \ensuremath{t_{s}=\frac{\xi_{0s}}{\theta_{s}}} and \ensuremath{\gamma_{s}=\frac{E_{0s}}{\theta_{s}}}}.
\end{equation}
In the above equations, the subscripts "p" and "s" symbolize the terms associated with pressure or compressional waves and shear waves respectively. Also, the material exponents $m$ and $n$ in Buckingham\citep{Buckingham2000} are reflected as $\gamma_s$ and $\gamma_p$ respectively in our calculations. This is done to avoid potential conflict with the symbols reserved for the framework of fractional calculus in Sec. III.

Buckingham then applies the Navier-Stokes equation to study the medium macroscopically resulting in the following two equations (see, Eq. (52) and (53) in Buckingham\citep{Buckingham2000}):
\begin{equation}\label{eq07}
\nabla^{2}\Psi-\frac{1}{c_{0}^{2}}\frac{\partial^{2}\Psi}{\partial t^{2}}+\frac{\lambda_{p}}{\rho_{0}c_{0}^{2}}\frac{\partial}{\partial t}\nabla^{2}\left[h_{p}\left(t\right)\ast\Psi\right]+\frac{4}{3}\frac{\eta_{s}}{\rho_{0}c_{0}^{2}}\frac{\partial}{\partial t}\nabla^{2}\left[h_{s}\left(t\right)\ast\Psi\right]=0
\end{equation}
and
\begin{equation}\label{eq08}
\frac{\eta_{s}}{\rho_{0}}\nabla^{2}\left[h_{s}\left(t\right)\ast A\right]-\frac{\partial A}{\partial t}=0
\end{equation}
where, $\rho_{0}$ is the bulk density of the material, $c_{0}=\sqrt{\frac{E_{0}}{\rho_{0}}}$ is the lossless phase velocity at zero frequency, and $\lambda_{p}$ and $\eta_{s}$ are stress relaxation coefficients corresponding to compressional waves and shear waves respectively. The terms appearing in the convolution terms of the two equations Eqs. (\ref{eq07}) and (\ref{eq08}) are related to the velocity vector $v$ as (see, Eq. (51) in Buckingham\citep{Buckingham2000}):
\begin{equation}\label{eq09}
v=\nabla\Psi+\nabla\times A.
\end{equation}
The above expression suggests that $\Psi$ and $A$ correspond to the wave displacement field for compressional and shear waves respectively.

\medskip

\section{FRAMEWORK OF FRACTIONAL CALCULUS}

Fractional calculus though as old as the classical Newtonian calculus was rediscovered only lately by Caputo\citep{Caputo1967} in $1967$ to explain dissipative mechanisms of materials often encountered in seismology and metallurgy.\citep{Caputo1971a}${}^,$\citep{Caputo1971b} Besides offering a mathematical extension to the regular integer-order derivatives, fractional derivatives facilitate modelling of materials characterized by spatial and/or temporal memory kernels with arbitrary-order exponents. One of the most used forms of the fractional derivative is by Caputo and is defined as the convolution of the power-law memory kernel $\Phi_{m}\left(t\right)$ with the ordinary derivative:\citep{Mainardi2010}
\begin{equation}\label{eq10}
\frac{d^{m}}{dt^{m}}f\left(t\right)\equiv{}_{0}D_{t}^{m}f\left(t\right)\triangleq \Phi_{m}\left(t\right)\ast \left(\frac{d^{n}}{d\tau^{n}}f\left(\tau\right)\right),
\end{equation}
where,
\begin{equation}\label{eq11}
\Phi_{m}\left(t\right)=\frac{t^{n-m-1}}{\varGamma\left(n-m\right)}.
\end{equation}
Using Eqs. (\ref{eq10}) and (\ref{eq11}), we have
\begin{equation}\label{eq12}
\frac{d^{m}}{dt^{m}}f\left(t\right)=\frac{1}{\varGamma\left(n-m\right)}\int\limits _{0}^{t}\frac{1}{\left(t-\tau\right)^{m+1-n}}\left(\frac{d^{n}}{d\tau^{n}}f\left(\tau\right)\right)d\tau.\end{equation}
Here $f\left(t\right)$ is a well behaved, causal, continuous function, $n$ is a positive integer, and the real-valued fractional-order $m\in\left(n-1,n\right)$. Also, $\varGamma\left(\cdot\right)$ is the Euler Gamma function defined for a complex variable $z$ as:
\begin{equation}\label{eq13}
\varGamma\left(z\right)=\int\limits _{0}^{\infty}x^{z-1}e^{-x}dx, \text{ } \Re\left(z\right)>0.
\end{equation}
From Eqs. (\ref{eq10})--(\ref{eq12}) it can be seen that the power-law memory kernel is built into the fabric of fractional calculus.

Substituting $n=0$ and $m$ by $-m$ in Eq. (\ref{eq12}) we obtain the corresponding expression for fractional integral as,
\begin{equation}\label{eq14}
\frac{d^{-m}}{dt^{-m}}f\left(t\right)\equiv{}_{0}I_{t}^{m}f\left(t\right)=\frac{1}{\varGamma\left(m\right)}\int\limits _{0}^{t}\frac{f\left(\tau\right)}{\left(t-\tau\right)^{1-m}}d\tau.\end{equation}
Since the Fourier transform of $\Phi_{m}\left(t\right)$ expressed by Eq. (\ref{eq11}) is a power-law in the frequency domain, it may be even easier to see the extension from the regular integer-order derivatives to the fractional-order derivatives from:
\begin{equation}\label{eq15}
\frac{d^{m}}{dt^{m}}f\left(t\right)\overset{\mathcal{F}}{\Longrightarrow}\left(i\omega\right)^{m}\hat{f}\left(\omega\right),
\end{equation}
where the spatio-temporal Fourier transform is defined as
\begin{equation}\label{eq16}
\mathcal{F}\left[f\left(x,t\right)\right]=\hat{f}\left(k,\omega\right)\triangleq\int\limits _{-\infty}^{\infty}\int\limits _{-\infty}^{\infty}f\left(x,t\right)e^{i\left(kx-\omega t\right)}dx\mbox{ }dt,
\end{equation}
where, $i=\sqrt{-1}$ is the imaginary number, $\omega$ is the temporal angular frequency and $k$ is the corresponding spatial frequency. It should be noted that the choice of positive and negative sign of the kernel gives two definitions of the Fourier transform (see, Appendix A\citep{Holm2014}).

\medskip

\subsection{Kelvin-Voigt fractional derivative wave equation}

Since the Kelvin-Voigt fractional derivative model is central to this paper, we present an illustration of its mechanical equivalent in Fig. 2.

\begin{figure}[H]
\centering
\hspace*{0cm}
\includegraphics[scale=1.3]{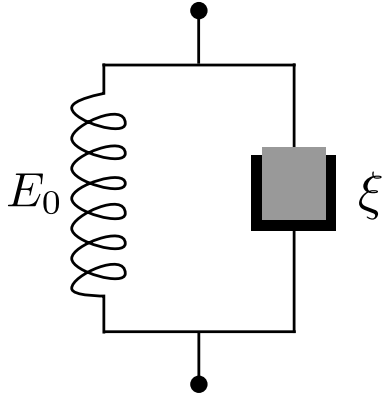}
\caption{(Color online) A mechanical equivalent sketch of the Kelvin-Voigt fractional derivative model comprising a parallel combination of a Hookean spring $E_{0}$ and a fractional viscous dashpot $\xi$.}
\end{figure}
The constitutive stress-strain relation of the model is given as\citep{Mainardi2010}
\begin{equation}\label{eq17}
\chi\left(t\right)=E_{0}\left[\varepsilon\left(t\right)+\tau_{\chi}^{m}\frac{d^{m}\varepsilon\left(t\right)}{dt^{m}}\right],
\end{equation}
where, the symbols represent the same physical quantities as in the case of Buckingham's GS model discussed in Sec. II and the last term with the fractional-order derivative corresponds to a "fractional" dashpot. The strain when expressed as a time-fractional derivative implies that the material has a long-term memory and remembers its past deformations via a fading memory weighted by a power-law function. The additional parameter $\tau_{\chi}$ is the characteristic retardation time of the material which gives a measure of the time taken for the creep strain to accumulate. A detailed analysis of the Kelvin-Voigt fractional derivative model is available.\citep{Holm2010}${}^-$\citep{Holm2013} The fractional-order $m$ is usually in the range from $0$ to $2$ , where $m=1$ corresponds to the standard viscoelastic case. The constitutive equation when combined with the laws of conservation of mass and momentum which are given as:
\begin{equation}\label{eq18}
\varepsilon\left(t\right)=\frac{\partial}{\partial x}u\left(x,t\right),
\end{equation}
and
\begin{equation}\label{eq19}
\nabla\chi\left(t\right)=\rho_{0}\frac{\partial^{2}}{\partial t^{2}}u\left(x,t\right)
\end{equation}
yields the following wave equation,
\begin{equation}\label{eq20}
\nabla^{2}u-\frac{1}{c_{0}^{2}}\frac{\partial^{2}u}{\partial t^{2}}+\tau_{\chi}^{m}\frac{d^{m}}{dt^{m}}\nabla^{2}u=0.
\end{equation}
Here, $u\left(x,t\right)=e^{i\left(\omega t-kx\right)}$ is the displacement of unit amplitude plane wave in space $x$ and time $t$ and the identity $\nabla^{2}$ is the Laplacian. We reserve the discussion of the dispersive characteristics from the Kelvin-Voigt fractional derivative wave equation until next section where we will encounter this equation again.

\medskip

\subsection{Time-fractional diffusion-wave equation}

The spring in the Kelvin-Voigt model when removed gives the following constitutive relation
\begin{equation}\label{eq21}
\chi\left(t\right)=E_{0}\tau_{\chi}^{m}\frac{d^{m}\varepsilon\left(t\right)}{dt^{m}}.
\end{equation}
Eq. (\ref{eq21}) when coupled with Eqs. (\ref{eq18}) and (\ref{eq19}) yields a diffusion-wave equation given as:
\begin{equation}\label{eq22}
\frac{\partial^{2-m}u}{\partial t^{2-m}}=D\frac{\partial^{2}u}{\partial x^{2}}
\end{equation}
where,  $D=\frac{E_{o}\tau_{\chi}^{m}}{\rho_{0}}>0$ is a constant with the dimension of $L^{2}T^{m-2}$.

In the limit as $m\rightarrow0$, the constitutive Eq. (\ref{eq21}) corresponds to that of an ideal spring which is also witnessed from Eq. (\ref{eq22}) as it approaches a lossless wave equation with $\sqrt{D}$ as the wave velocity. On the other hand, in the limit as $m\rightarrow1$, the spring transforms into a pure damper, Eq. (\ref{eq22}) then approaches a standard diffusion equation with $D$ as the diffusion coefficient. For arbitrary values of $m\in\left[0,1\right]$, Eq. (\ref{eq22}) predicts an interpolation between diffusion and wave like behavior.\citep{Agrawal2002}${}^,$\citep{Ray2007} However, if $m\in\left[1,2\right]$, Eq. (\ref{eq22}) would then represent a sub-diffusion process. 

\medskip

\section{FROM GS MODEL TO FRACTIONAL WAVE EQUATIONS}

Following the argument that the sliding between micro-asperities is indistinguishable for compressional and shear stress relaxations (see, Eqs. (31) and (32) in Buckingham\citep{Buckingham2000}), we drop the notations of "p" and "s" from Eqs. (\ref{eq04})--(\ref{eq06}) such that
\begin{equation}\label{eq23}
t_{p}=t_{s}=\tau,
\end{equation}
\begin{equation}\label{eq24}
\gamma_{p}=\gamma_{s}=\gamma,
\end{equation}
\begin{equation}\label{eq25}
\mbox{and thus, } h_{p}=h_{s}=h.
\end{equation}
Here, we would like to mention that the MIRFs for the compressional wave and shear wave are not same. This is because the compressional viscoelastic time constant $t_{p}$ and shear viscoelastic time constant $t_{s}$ are not equal, but rather $\frac{t_{s}}{t_{p}}\approx10$. This correction modified the VGS model to become the VGS($\lambda$) model which gave a better fit to the shear wave dispersion measurements.\citep{Buckingham2007}${}^-$\citep{Buckingham2010}

Using  Eqs. (\ref{eq23})--(\ref{eq25}) we rewrite the expression of MIRF from Eq. (\ref{eq05}) as,
\begin{equation}\label{eq26}
h\left(t\right)=\tau^{-1}\left(1+\frac{t}{\tau}\right)^{-\gamma}\text{, where \ensuremath{\tau=\frac{\xi_{0}}{\theta}} and \ensuremath{\gamma=\frac{E_{0}}{\theta}}}.
\end{equation}
We stress that Eq. (\ref{eq26}) is actually a stretched asymptotic power-law which demonstrates long-time inverse power-law behaviour and therefore can be approximated as the Nutting law: 
\begin{equation}\label{eq27}
h\left(t\right)\sim\tau^{-1}\left(\frac{t}{\tau}\right)^{-\gamma}
\end{equation}
Since, the approximation is better the larger $t$ is relative to $\tau$, the VGS($\lambda$) model actually implies that the approximation is slightly better for compressional waves than for shear waves. Further, Eq. (\ref{eq27}) is equivalent to the Kohlrausch-Williams-Watts (KWW) function which is given as,\citep{Das2010}
\begin{equation}\label{eq28}
h\left(t\right)=\tau^{-1}e^{-\left(\frac{t}{\tau}\right)^{\gamma}}\mbox{, where }\gamma\in\left(0,1\right).
\end{equation}
The equivalence of the MIRF from the GS model and KWW function suggests that the strain-hardening mechanism is essentially a non-Debye or non-exponential relaxation process. Therefore, we infer that the relaxation from strain-hardened grain-shearing in fluid-saturated unconsolidated sediments is inherently non-Markovian. In other words, the material possesses temporal memory which paves the way for mapping the GS model into the fractional framework. This is also witnessed from the similarity of Eq. (\ref{eq27}) with the memory kernel of Eq. (\ref{eq11}).

\medskip

\subsection{Compressional wave equation}

Using Eq. (\ref{eq24}) the last two terms in the compressional wave Eq. (\ref{eq07}) can be merged together as,
\begin{equation}\label{eq29}
\nabla^{2}\Psi-\frac{1}{c_{0}^{2}}\frac{\partial^{2}\Psi}{\partial t^{2}}+\ensuremath{\left(\frac{\lambda_{p}}{\rho_{0}c_{0}^{2}}+\frac{4}{3}\frac{\eta_{s}}{\rho_{0}c_{0}^{2}}\right)}\frac{\partial}{\partial t}\nabla^{2}\left[h\left(t\right)\ast\Psi\right]=0
\end{equation}
Substituting Eq. (\ref{eq27}) in Eq. (\ref{eq29}) we have,
\begin{equation}\label{eq30}
\nabla^{2}\Psi-\frac{1}{c_{0}^{2}}\frac{\partial^{2}\Psi}{\partial t^{2}}+\ensuremath{\left(\frac{\lambda_{p}}{\rho_{0}c_{0}^{2}}+\frac{4}{3}\frac{\eta_{s}}{\rho_{0}c_{0}^{2}}\right)}\tau^{\gamma-1}\frac{\partial}{\partial t}\nabla^{2}\left[ t^{-\gamma}\ast\Psi\right]=0.
\end{equation}
Manipulating the last convolution term as
\begin{equation}\label{eq31}
\left[ t^{-\gamma}\ast\Psi\right]=\varGamma\left(1-\gamma\right)\left[\Psi\ast\frac{t^{-\gamma}}{\varGamma\left(1-\gamma\right)}\right]=\varGamma\left(1-\gamma\right)\left[\frac{d^{1}}{dt^{1}}\left\{ \frac{d^{-1}}{dt^{-1}}\Psi\right\} \ast\frac{t^{-\gamma}}{\varGamma\left(1-\gamma\right)}\right],
\end{equation}
and then comparing Eq. (\ref{eq31}) with  Eq. (\ref{eq10}), we identify $f\left(\tau\right)=\frac{d^{-1}}{dt^{-1}}\Psi$, $n=1$ and $m=\gamma\Rightarrow \gamma\in\left(0, 1\right)$ which is in accordance to Buckingham.\citep{Buckingham2007} The term with negative fractional-order time derivative in Eq. (\ref{eq31}) is equivalent to the fractional integration given by Eq. (\ref{eq14}). Using Eqs. (\ref{eq10}), (\ref{eq11}) and (\ref{eq31}), the fractional-order derivative equivalent of the convolution term in Eq. (\ref{eq30}) can be written as,
\begin{equation}\label{eq32}
\left[ t^{-\gamma}\ast\Psi\right]=\varGamma\left(1-\gamma\right)\frac{d^{\gamma-1}}{dt^{\gamma-1}}\Psi.
\end{equation}
Substituting Eq. (\ref{eq32}) back in Eq. (\ref{eq30}) and rearranging the terms we get,
\begin{equation}\label{eq33}
\nabla^{2}\Psi-\frac{1}{c_{0}^{2}}\frac{\partial^{2}\Psi}{\partial t^{2}}+\varGamma\left(1-\gamma\right)\ensuremath{\left(\frac{\lambda_{p}}{\rho_{0}c_{0}^{2}}+\frac{4}{3}\frac{\eta_{s}}{\rho_{0}c_{0}^{2}}\right)\tau^{\gamma-1}}\frac{d^{\gamma}}{dt^{\gamma}}\nabla^{2}\Psi=0.
\end{equation}
We find that Eq. (\ref{eq33}) is equivalent to the Kelvin-Voigt fractional derivative wave Eq. (\ref{eq20}), where
\begin{equation}\label{eq34}
\tau_{\chi}^{\gamma}=\varGamma\left(1-\gamma\right)\ensuremath{\left(\frac{\lambda_{p}}{\rho_{0}c_{0}^{2}}+\frac{4}{3}\frac{\eta_{s}}{\rho_{0}c_{0}^{2}}\right)\tau^{\gamma-1}}.
\end{equation}
The dimensional consistency of Eq. (\ref{eq34}) suggests the validity of  Eq. (\ref{eq33}) and hence the mapping of compressional wave Eq. (\ref{eq07}) of the GS model into the fractional framework. Also, from Eq. (\ref{eq34}) we obtain the relationship between the characteristic relaxation time constant of a material with its geo-acoustic parameters. Further, we stress that the fractional-order $\gamma$ which characterises the constitutive stress-strain Eq. (\ref{eq17}) and wave Eq. (\ref{eq20}) of Kelvin-Voigt fractional derivative model gains a physical interpretation as mentioned in Eq. (\ref{eq26}). The order $\gamma=\frac{E_{0}}{\theta}$ gives a measure of the interplay between the elastic and viscoleastic properties of the material. 

In the limiting case of maximum strain-hardening, i.e. if $\theta\rightarrow\infty\Rightarrow\gamma\rightarrow0$, Eq. (\ref{eq33}) approaches the familiar lossless wave equation. Physically, it implies that the intergranular sliding has stopped which is possible if grains are locked against each other. Such a situation is plausible at-least locally among the participating grains if the pore-fluid is completely squeezed out as a result of the intergranular sliding. In an ideal condition, the initially assumed unconsolidated granular material would then effectively transform into a compact solid and therefore any possible energy dissipation in the viscous pore-fluid would be ruled out. On the other hand, as strain-hardening decreases, i.e. $\gamma$ increases, grain-shearing is facilitated and attenuation rises. Besides, in such cases varying degrees of flow of the pore-fluid in between the grains cannot be neglected. In the limit as $\gamma\rightarrow1$, the wave Eq. (\ref{eq33}) approaches the classical viscous wave equation. 

The dispersive behaviour from the Kelvin-Voigt fractional derivative wave Eq. (\ref{eq20}) has already been studied in detail.\citep{Holm2010}${}^-$\citep{Holm2013} However for completeness we repeat the necessary mathematical framework which will be utilized in the next subsection when we analyze the shear wave equation.

For modeling of dispersive properties in a material, we assume the wave propagation vector $k$ to be complex such that
\begin{equation}\label{eq35}
k=\beta_{k}-i\alpha_{k}, \text{ } \beta_{k}\geq0 \text{ and } \alpha_{k}\geq0,
\end{equation}
where, $\beta_{k}$ is the wave velocity vector and $\alpha_{k}$ is the wave attenuation vector. Both $\beta_{k}$ and $\alpha_{k}$ are functions of $\omega$, and also related to each other by the Kramers-Kronig relations due to causality.\citep{Szabo1995} Since, the Kramers-Kronig relations are fundamentally the Hilbert transform pair, it couples attenuation of the wave with its velocity. 

We further assume the initial wave disturbance in the form of an unit amplitude, one-dimensional, plane wave given as $\Psi\left(x,t\right)=e^{i\left(\omega t-kx\right)}$. Then using Eq. (\ref{eq35}), we obtain the expression for a propagating wave field in a lossy medium as,
\begin{equation}\label{eq36}
\Psi\left(x,t\right)=e^{-\alpha_{k}x}e^{i\left(\omega t-\beta_{k}x\right)}.
\end{equation}
The phase and group velocities of the wave are given by the real parts of $k$ as $c_{p}\left(\omega\right)=\omega/\beta_{k}$ and $c_{g}\left(\omega\right)=d\omega/d\beta_{k}$. On Fourier-transforming the wave Eq. (\ref{eq33}) and using Eq. (\ref{eq15}), we obtain the dispersion relation as
\begin{equation}\label{eq37}
k=\frac{\omega}{c_{0}}\sqrt{\frac{1}{1+\left(i\omega\tau_{\chi}\right)^{\gamma}}}
\end{equation}
where, $\tau^{\gamma}_{\chi}$ is given by Eq. (\ref{eq34}). As can be seen in the limit as $\gamma\rightarrow0$, Eq. (\ref{eq37}) approaches the lossless dispersion relation. On the contrary, in the limit as $\gamma\rightarrow1$, Eq. (\ref{eq37}) approaches the classical damped viscous-wave dispersion relation. 

We set the following numerical values for plotting purpose: $c_{0}=1$, $\tau=1$ and $\tau_{\chi}^{\gamma}=\varGamma\left(1-\gamma\right)$. The phase velocity dispersion curve and attenuation curve for different values of $\gamma$ are shown in Fig. 3 and Fig. 4 respectively.
\begin{figure}[H]
\centering
\hspace*{0cm}
\includegraphics[scale=0.39]{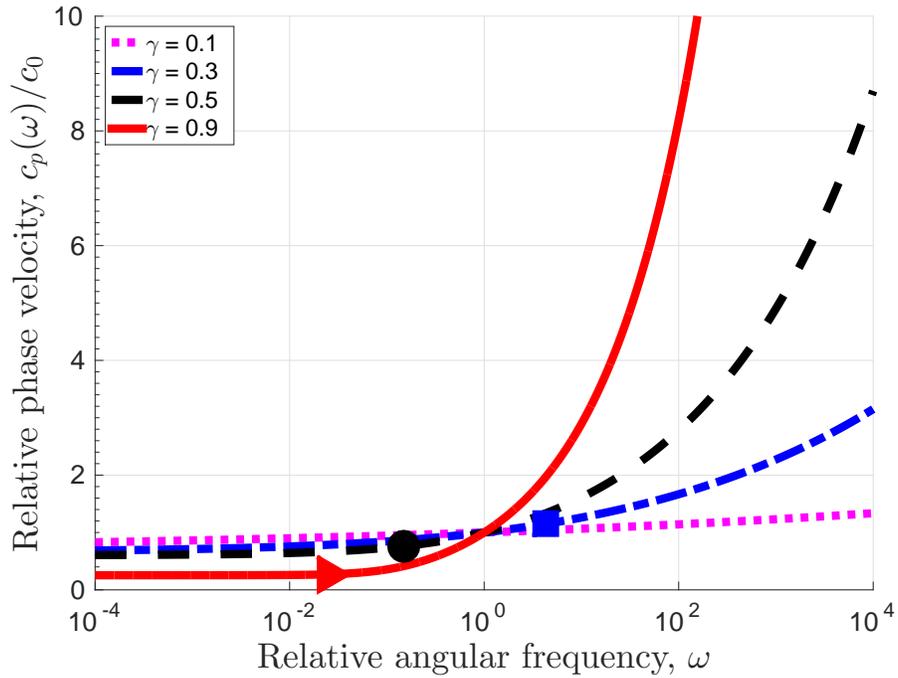}
\caption{(Color online) Frequency-dependent phase velocity dispersion
for the compressional waves from the GS model. The fractional derivative order $\gamma$ has values 0.1 (dotted line), 0.3 (dash-dot line), 0.5 (dashed line) and 0.9 (solid line). The markers; square for $\gamma=0.3$, circle for $\gamma=0.5$ and triangle for $\gamma=0.9$ represent the upper
cut-off of the dispersion curve corresponding to a penetration depth
of one wavelength. The marker for $\gamma=0.1$ lies outside the given
frequency range. Each curve is normalized to its value at $\omega=1\mbox{ } rad\mbox{ }s^{-1}$.
 \smallskip}
\end{figure}

\begin{figure}[H]
\centering
\hspace*{0cm}
\includegraphics[scale=0.39]{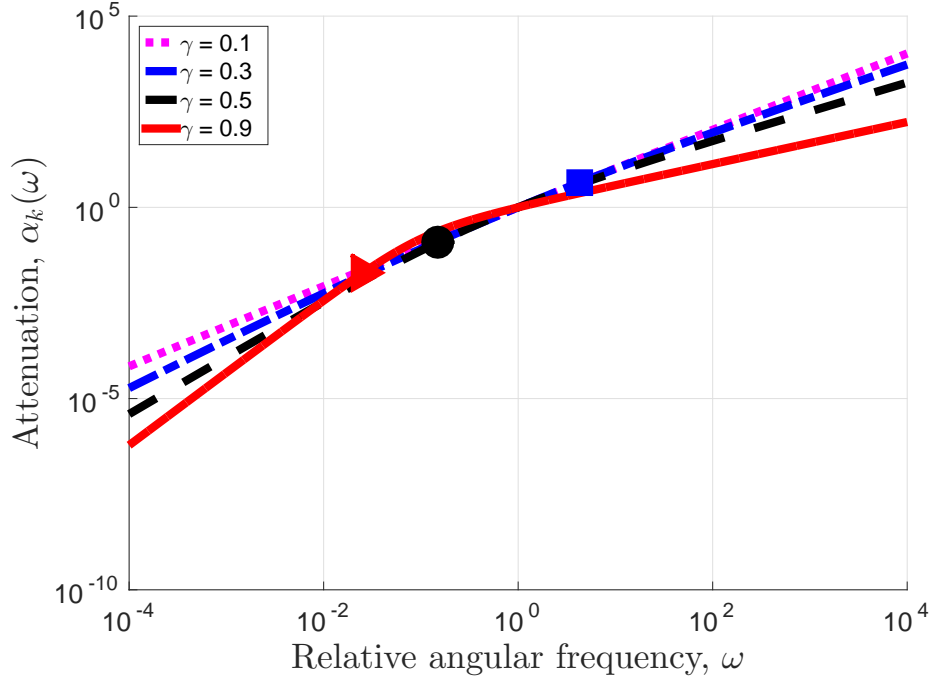}
\caption{(Color online) Frequency-dependent wave attenuation for the compressional waves from the GS model. The fractional derivative order $\gamma$ has values 0.1 (dotted line), 0.3 (dash-dot line), 0.5 (dashed line) and 0.9 (solidline). The markers; square for $\gamma=0.3$, circle for $\gamma=0.5$ and triangle for $\gamma=0.9$ represent the upper cut-off of the attenuation curve corresponding to a penetration depth of one wavelength. The marker for $\gamma=0.1$ lies outside the given frequency range. Each curve is normalized to its value at $\omega=1\mbox{ } rad\mbox{ }s^{-1}$.}
\end{figure}

The asymptotic behavior of the dispersion plots remains the same as discussed in Holm and Sinkus.\citep{Holm2010} Summarizing them here:
\begin{equation}\label{eq38}
\alpha_{k}\left(\omega\right)\begin{cases}
\propto\omega^{1+\gamma}, & \left(\omega\tau_{\chi}\right)^{\gamma}\ll1\\
\propto\omega^{1-\frac{\gamma}{2}}, & 1\ll\left(\omega\tau_{\chi}\right)^{\gamma}
\end{cases}
\end{equation}
and, 
\begin{equation}\label{eq39}
c_{p}\left(\omega\right)\begin{cases}
=c_{0}, & \left(\omega\tau_{\chi}\right)^{\gamma}\ll1\\
\propto\omega^{\frac{\gamma}{2}}, & 1\ll\left(\omega\tau_{\chi}\right)^{\gamma}.
\end{cases}
\end{equation}
The conditions $\left(\omega\tau_{\chi}\right)^{\gamma}\ll1$ and $1\ll\left(\omega\tau_{\chi}\right)^{\gamma}$ mentioned in above Eqs. (\ref{eq38}) and (\ref{eq39}) characterize the low frequency and high frequency regimes respectively.

As illustrated in Fig. 3 and also seen from the asymptotic behavior expressed by Eq. (\ref{eq39}), for larger values of $\gamma$, as $\omega\rightarrow\infty$, phase velocity $c_{p}\rightarrow\infty$. This apparently looks like a violation of causality however it is not, since as $\omega\rightarrow\infty$, attenuation $\alpha_{k}\rightarrow\infty$ (see, Fig. 4). Moreover, it can be seen from Eqs. (\ref{eq38}) and (\ref{eq39}) that the rate of increase of attenuation is comparatively higher than that of the phase velocity for same frequency values. This aspect of the dispersive behavior can be better explained by employing the notion of penetration depth or skin depth which is often used in electromagnetic studies.\citep{Ziomek1994} The penetration depth $\delta_{p}$ of the propagating wave is defined as the distance traversed by the wave before its amplitude falls by a factory of $\frac{1}{e}\approx36.7\mbox{ }\%$, or intensity becomes $\frac{1}{e^{2}}\approx13.5\mbox{ }\%$ of its maximum value. Imposing this condition on Eq. (\ref{eq36}) gives,
\begin{equation}\label{eq40}
\delta_{p}\left(\omega\right)=\frac{1}{\alpha_{k}\left(\omega\right)}.
\end{equation}
The penetration depth $\delta_{p}$ can be better quantified in number of wavecycles as,
\begin{equation}\label{eq41}
\delta_{p}\left(\omega\right)=\frac{\omega}{2\pi\alpha_{k}\left(\omega\right)c_{p}\left(\omega\right)}\mbox{ (in number of wavelengths).} 
\end{equation}
Using Eqs.  (\ref{eq38}), (\ref{eq39}) and (\ref{eq41}) we obtain the asymptotic behavior of the penetration depth in low frequency and high frequency regimes as:
\begin{equation}\label{eq42}
\delta_{p}\left(\omega\right)\begin{cases}
\propto\omega^{-\gamma}, & \left(\omega\tau_{\chi}\right)^{\gamma}\ll1\\
=\text{constant}, & 1\ll\left(\omega\tau_{\chi}\right)^{\gamma}.
\end{cases}
\end{equation}
As illustrated in Fig. 5, the value of the penetration depth is greater for low frequencies. With increase in frequency, the penetration depth obeying power-law given by Eq (\ref{eq42}) falls and approaches a constant value for high frequencies. In the case of viscous media, i.e. for larger values of $\gamma$, the penetration depth falls even more rapidly and is reduced to less than a single wavelength in the high frequency regime. Physically, it implies that the wave enters the evanescent mode where its oscillatory motion ceases to exist and the wave finally decays.

\begin{figure}[H]
\centering
\hspace*{0cm}
\includegraphics[scale=0.39]{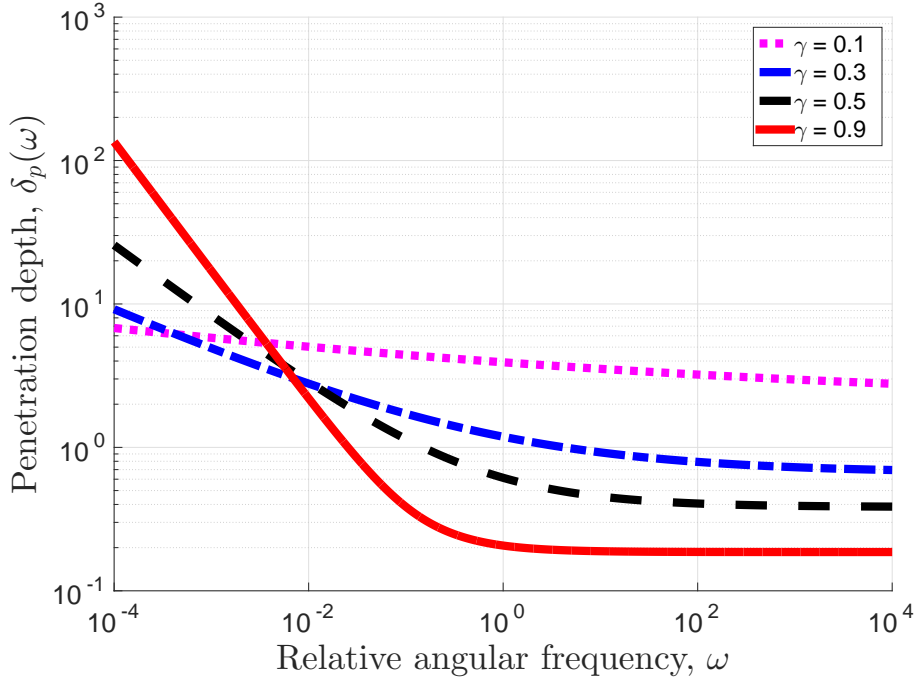}
\caption{(Color online) Frequency-dependent penetration depth (in
number of wavelengths) for the compressional waves from the
GS model. The fractional derivative order $\gamma$ has values 0.1 (dotted line), 0.3 (dash-dot line), 0.5 (dashed line) and 0.9 (solid line).}
\end{figure}

To illustrate this idea even better, we choose a threshold distance of a single wavelength. The upper cut-off of the dispersion curves which allows the wave to travel a minimum distance equal to its wavelength is depicted by markers in Fig. 3 and Fig. 4. As seen from Fig. 5, the wave can travel significantly larger distance for small value of $\gamma=0.1$ in the entire frequency regime before it undergoes exponential damping. This is also reflected by the absence of a marker corresponding to the value of $\gamma=0.1$ in Fig. 3 and Fig. 4, since it lies outside the given frequency range. Also, for such small values of $\gamma$ the phase velocity does not rises as rapidly as for other larger values of $\gamma$. This observation regarding small values of $\gamma$ are the most interesting ones because they are used by Buckingham for the curve-fitting of experimental data with near-linear power-laws in frequency, i.e. exponents with small $\gamma$ values.\citep{Buckingham2007}

\medskip

\subsection{Shear wave equation}

Replacing the wave displacement field $\Psi$ by $A$ in Eq. (\ref{eq32}) and then substituting it along with Eq. (\ref{eq26}) in Eq. (\ref{eq08}), we get
\begin{equation}\label{eq43}
\varGamma\left(1-\gamma\right)\frac{\eta_{s}}{\rho_{0}}\tau{}^{\gamma-1}\nabla^{2}\left[\frac{d^{\gamma-1}}{dt^{\gamma-1}}A\right]-\frac{\partial A}{\partial t}=0.
\end{equation}
On further simplification Eq. (\ref{eq43}) becomes
\begin{equation}\label{eq44}
\frac{\partial^{2-\gamma}A}{\partial t^{2-\gamma}}=\varGamma\left(1-\gamma\right)\frac{\eta_{s}}{\rho_{0}}\tau^{\gamma-1}\nabla^{2}A.
\end{equation}
Comparing Eq. (\ref{eq44}) with Eq. (\ref{eq22}), we find that the shear wave equation from the GS model is equivalent to a time-fractional diffusion-wave equation. Further, the diffusion coefficient of the process is identified as
\begin{equation}\label{eq45}
D=\frac{\eta_{s}}{\rho_{0}}\tau^{\gamma-1}\varGamma\left(1-\gamma\right).
\end{equation}
Then Eq. (\ref{eq44}) attains its final form:
\begin{equation}\label{eq46}
\frac{\partial^{2-\gamma}A}{\partial t^{2-\gamma}}=D\nabla^{2}A.
\end{equation}
Since $0<\gamma<1$, the fractional-order in Eq. (\ref{eq46}) is, $1<\left(2-\gamma\right)<2$. The time domain solutions of the equation suggest that as $\gamma$ decreases from $1$ to $0$, the phenomenon of diffusion transforms into a lossless wave propagation.\citep{Agrawal2002}${}^,$\citep{Ray2007} 
Following the same approach as in the case of compressional wave equation, we can examine the dispersion characteristic of the fractional diffusion-wave equation. We Fourier transform Eq. (\ref{eq46}) to obtain the following dispersion relation,
\begin{equation}\label{eq47}
k=\frac{i^{-\frac{\gamma}{2}}}{\sqrt{D}}\omega^{1-\frac{\gamma}{2}}.
\end{equation}
Further, using Eq. (\ref{eq35}) in Eq. (\ref{eq47}) we have,
\begin{equation}\label{eq48}
\beta_{k}-i\alpha_{k}=\frac{\omega^{1-\frac{\gamma}{2}}}{\sqrt{D}}\left\{ \cos\left(\gamma\frac{\pi}{4}\right)-i\mbox{ }\sin\left(\gamma\frac{\pi}{4}\right)\right\}.
\end{equation}
Comparing the real and imaginary parts of Eq. (\ref{eq48}), we get:
\begin{equation}\label{eq49}
\beta_{k}\left(\omega\right)=\frac{\omega^{1-\frac{\gamma}{2}}}{\sqrt{D}}\cos\left(\gamma\frac{\pi}{4}\right)
\end{equation}
and,
\begin{equation}\label{eq50}
\alpha_{k}\left(\omega\right)=\frac{\omega^{1-\frac{\gamma}{2}}}{\sqrt{D}} \sin\left(\gamma\frac{\pi}{4}\right).
\end{equation}
The phase velocity is then expressed as:
\begin{equation}\label{eq51}
c_{p}\left(\omega\right)=\frac{\omega}{\beta_{k}\left(\omega\right)}=\sqrt{D}\frac{\omega^{\frac{\gamma}{2}}}{\cos\left(\gamma\frac{\pi}{4}\right)}.
\end{equation}
And, the penetration depth is,
\begin{equation}\label{eq52}
\delta_{p}\left(\omega\right)=\frac{\omega}{2\pi\alpha_{k}\left(\omega\right)c_{p}\left(\omega\right)}=\frac{1}{2\pi}\cot\left(\gamma\frac{\pi}{4}\right) \mbox{(in number of wavelengths)}.
\end{equation}
In the limit as $\gamma\rightarrow0$, $\delta_{p}\rightarrow\infty$, which physically implies the classical lossless wave propagation. On the other hand, as $\gamma\rightarrow1$, $\delta_{p}\rightarrow\frac{1}{2\pi}<1$; this corresponds to the characteristic evanescent wave solution expected from the standard diffusion equation.

We set the following numerical values for plotting purpose: $\tau=1$ and $D=\varGamma\left(1-\gamma\right)$ to include the $\gamma$ dependency on dispersion. The phase velocity dispersion curve and attenuation curve for different values of $\gamma$ are illustrated in Fig. 6 and Fig. 7 respectively. As in the case of the compressional wave in the GS model, causality is ensured for the shear wave as well. For example, if $\gamma\rightarrow1$, then from Eq. (\ref{eq51}) as $\omega\rightarrow\infty$, $c_{p}\rightarrow\infty$, however penetration depth given by Eq. (\ref{eq52}) is limited to less than a single wavelength.

\begin{figure}[H]
\centering
\hspace*{0cm}
\includegraphics[scale=0.39]{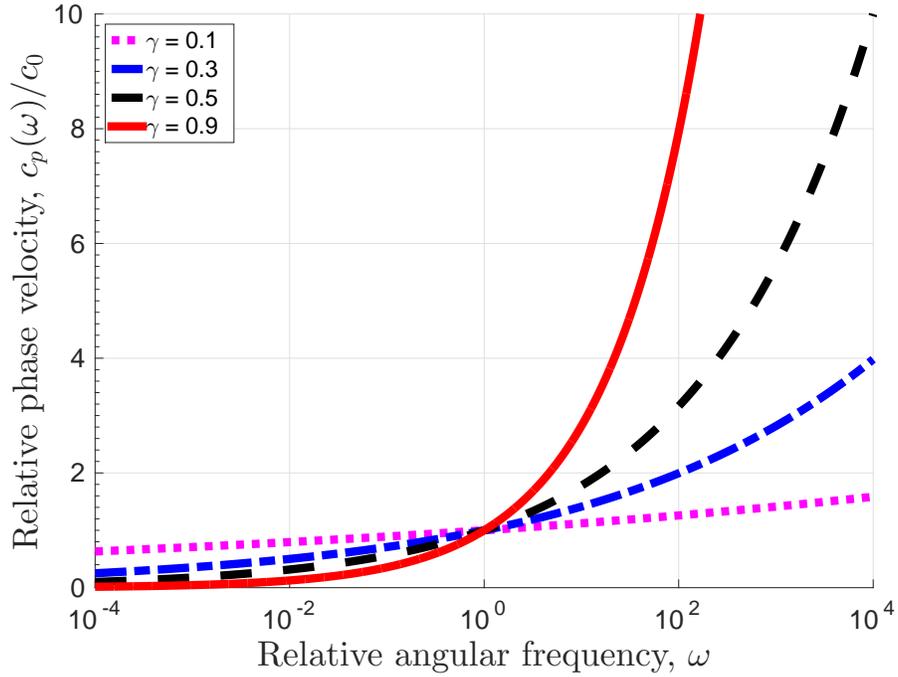}
\caption{(Color online) Frequency-dependent phase velocity dispersion
for the shear waves from the GS model. The fractional derivative order $\gamma$ has values 0.1 (dotted line), 0.3 (dash-dot line), 0.5 (dashed line) and 0.9 (solid line). Each curve is normalized to its value at $\omega=1\mbox{ } rad\mbox{ }s^{-1}$.}
\end{figure}

\begin{figure}[H]
\centering
\hspace*{0cm}
\includegraphics[scale=0.39]{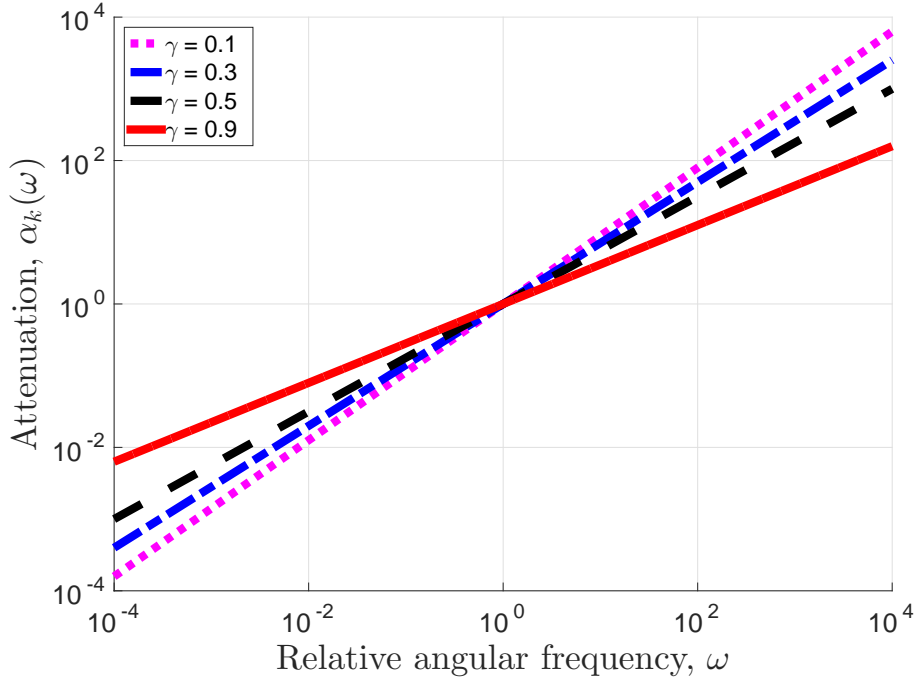}
\caption{(Color online) Frequency-dependent wave attenuation for the shear waves from the GS model. The fractional derivative order $\gamma$ has values 0.1 (dotted line), 0.3 (dash-dot line), 0.5 (dashed line) and 0.9 (solid line). Each curve is normalized to its value at $\omega=1\mbox{ } rad\mbox{ }s^{-1}$.}
\end{figure}

Unlike for the compressional wave equation, the shear wave propagation is characterized by a single power-law in the entire frequency regime. Comparing Eqs. (\ref{eq51}) and (\ref{eq39}), it can be seen that the phase velocity dispersion of the shear wave follows the same power-law as the compressional wave in the high frequency regime. Further, as seen from Eqs. (\ref{eq49}) and (\ref{eq50}), the components of the wave propagation vector, here given as the phase velocity vector and the attenuation vector follow the same power-law. The equality of the competing race between the vectors is witnessed in the frequency independent expression of the penetration depth given by Eq. (\ref{eq52}). 

\medskip

\section{DISCUSSION AND CONCLUSION}

The merit of the GS model lies in the fact that besides predicting the dispersion relation for compressional \& shear waves as functions of frequency, it also relates to the geo-acoustic parameters of the material, such as grain size, density, viscosity, porosity, permeability and over-burden pressure.\citep{Buckingham2005} Here, we would like to add a minor comment about the physical mechanism which generates time-dependent strain-hardening in the GS model. Buckingham attributes it to the drag force arising as a result of the sliding motion across the the pore-fluid present between the grains. However, we would like to bring to notice that the time-dependency of strain-hardening is a well established phenomenon in stick-slip processes occurring in dried granular material as well.\citep{Dieterich1978} As we understand, the origin of the grain-shearing could be due to the microscopic junctions formed as a result of the sliding of micro-asperities against each other. These micro-junctions have an inherent property of being time-dependent.\citep{Thogersen2014} But, this observation does not change the constitutive mathematical framework of GS model expressed by Eq. (\ref{eq01}). The time-dependent viscous dashpot $\xi\left(t\right)$ and hence the time-dependent strain-hardening in the GS model would then correspond to the inherent time-dependency of the intergranular sliding which is further enhanced by the "fluid-saturated" aspect of the grain sediments. Since the grains are saturated by the same pore-fluid, they gain viscoelastic characteristics and therefore the presence of $\xi\left(t\right)$ in Eq. (\ref{eq01}) is justified. The time-dependency arising due to the fluid-saturation of the grains is determined by the properties of the fluid as well as by the porosity and permeability of the grains. In the light of this finding, the term expressed by Eq. (24c) in Buckingham\citep{Buckingham2007} is actually the viscous drag force. A further extension of this work would be to study the predictions of the model if time-dependency of the drag force is taken into account.

One of the goals achieved through this work is that we have shown that the equations derived from the fundamental physical process of grain-shearing could lead to fractional-order wave equations. This may be the first result which directly connects the fractional framework to a process deeply rooted in physics. A bonus of this bridging is the physical significance of the fractional-order extracted from the GS model. An estimation of the order from the physical parameters of the material would now greatly reduce the ambiguities in the curve-fitting of experimental data with the predictions made using fractional-order wave equations. Besides, the framework may provide new insights and perspectives which could be otherwise difficult to predict from integer-order wave equations, such as diffusion-wave phenomena, or the underlying fractal geometry of the material.\citep{Nigmatullin1986}

Further, if the rotation of grains is included in the GS model, it can facilitate squeezing of the pore-fluid between the grains. Given the fact that the material is unconsolidated, for a given sample of randomly distributed coarse grains of varying shapes, sizes and orientations, there would be some configurations that would be more favorable to support the motion along the grain boundaries.\citep{Schumacher1941} It is reasonable to consider that the time-dependent compressional and tangential shearing would also cause the rotation of grains at frequent intervals. The rotation would then bring grains into positions which could either favor or oppose the intergranular sliding motion. This also gives possibilities for squeezing of the pore-fluid through the intergranular pores, i.e. migration from regions of high concentration to low concentration. Thus, it comes as no surprise that the shear wave equation from the GS model appears as a diffusion-wave equation in the framework of fractional calculus.\citep{Gifkins1968} 

The dispersion curves are often used to extract the viscoelastic parameters of the material by superimposing them with the master curves.\citep{Meidav1964} With increase in complexity of the material, there is a demand for a more flexible but yet robust mathematical framework for modelling the material behavior for many applications in the field of acoustics, medical ultrasound, seismology and geophysics. Fractional calculus is one of the candidates which could offer the required framework and utilities.

\medskip

 
 \setlength{\parindent}{0.7cm}

 \end{document}